\documentclass[preprint]{aastex}
\usepackage{natbib}
\usepackage{booktabs}
\usepackage{threeparttable}
\usepackage{longtable}
\usepackage{tabularx}
\usepackage{lineno}

\begin{document}

\title{Localizing quasi-periodic pulsations in hard X-ray, microwave and Ly$\alpha$ emissions of an X6.4 Flare}

\author{Dong~Li$^{1,2}$, Zhenxiang~Hong$^{3,1}$, Zhenyong~Hou$^4$, and Yang~Su$^1$}
\affil{$^1$Key Laboratory of Dark Matter and Space Astronomy, Purple Mountain Observatory, CAS, Nanjing 210023, PR China \\
       $^2$Yunnan Key Laboratory of the Solar physics and Space Science, Kunming 650216, PR China \\
       $^3$Zhejiang International Studies University, Hangzhou, 310023, PR China \\
       $^4$School of Earth and Space Sciences, Peking University, Beijing 100871, PR China
     }
     \altaffiltext{}{Correspondence should be sent to: lidong@pmo.ac.cn}
\begin{abstract}
We report the simultaneous observations of quasi-periodic pulsations
(QPPs) in wavelengths of hard X-ray (HXR), microwave, Ly$\alpha$,
and ultraviolet (UV) emissions during the impulsive phase of an X6.4
flare on 2024 February 22 (SOL2024-02-22T22:08). The X6.4 flare
shows three repetitive and successive pulsations in HXR and
microwave wavebands, and they have an extremely-large modulation
depth. The onset of flare QPPs is almost simultaneous with the start
of magnetic cancellation between positive and negative fields. The
wavelet power spectra suggest the presence of double periods, which
are centered at $\sim$200~s and $\sim$95~s, respectively. The
long-period QPP can also be detected in Ly$\alpha$ and UV wavebands
at the flare area, and it could be observed in the adjacent sunspot.
Our observations indicate that the flare QPPs are most likely
triggered by accelerated electrons that are associated with periodic
magnetic reconnections. The long period at $\sim$200~s is probably
modulated by the slow magnetoacoustic wave originating from the
neighboring sunspot, while the short period at $\sim$95~s could be
regarded as its second harmonic mode.
\end{abstract}


\section{Introduction}
Quasi-periodic pulsations (QPPs) are frequently detected in the time
series of solar flares, which are highly variable modulations to the
flare radiation, and could be applied to diagnose the plasma
fluctuations in the solar atmosphere \citep[e.g.,][references
therein]{Kupriyanova20,Zimovets21}. The key feature of flare QPPs is
repeatability, successiveness, and impulsiveness, which may play a
crucial role in diagnosing the intermittent and impulsive
energy-releasing process on the Sun \citep{Inglis23}. A
classic QPP event usually exhibits at least three successive
pulsations in the time series, there is no reason to discuss the
QPP feature if it shows just one or two pulsations, because this
might be a random occurrence \citep[e.g.,][]{Nakariakov19}. The
flare QPPs are often composed by a group of nonstationary
pulsations, termed as ``nonstationary QPPs'', and the pulsation
profile of this type is anharmonic and irregular
\citep[cf.][]{Nakariakov19}. The flare QPP was first discovered by
\cite{Parks69} in the X-ray channel. After then, they have been
observed in nearly all the electromagnetic wavebands, ranging from
radio/microwave emissions, through white light, ultraviolet (UV),
and extreme ultraviolet (EUV) wavelengths to soft/hard X-rays
(SXR/HXR) and even to $\gamma$-rays
\citep[e.g.,][]{Nakariakov10,Dennis17,Nakariakov18,Shen22,Li20a,Li22b,Mehta23,Motyk23,Yu24,Zhou24}.
The flare QPPs were also seen in the Ly$\alpha$ emission
\citep{Milligan17,Li21,Lu21}. The quasi-periods of flare QPPs were
observed in a wide range of timescales of minutes, seconds, or even
milliseconds
\citep[e.g.,][]{Sych09,Tan10,Kumar15,Karlicky20,Xu23,Zhao23,Millar24}.
It seems that the detected period is highly associated with the
observed waveband or the used instrument. Moreover, double or even
multiple periods were simultaneously observed in the time series of
the same flare, which could be explained as the fundamental and second
or multiple harmonics \citep{Inglis09,Lid22}. The QPPs feature could
also be found in the circular-ribbon flare, and this type of flare
often occurs in a complicated magnetic structure associated with a
fan-spine topology \citep[cf.][]{Torok09,Zhang24}, resulting in
various kinds of flare QPPs
\citep[e.g.,][]{Kashapova20,Ning22,Altyntsev22}.

Although the flare QPPs have been discovered more than fifty years
\citep[see,][]{Parks69}, it is still in debate which generation
mechanism should be responsible for triggering a specific QPP event,
mainly because that the current observational data could not provide
sufficient information to distinguish various mechanisms
\citep[e.g.,][]{Inglis23}. Assuming the presence of
magnetohydrodynamic (MHD) waves and periodic magnetic reconnection
on the Sun, more than fifteen types of mechanisms or models are
proposed to interpret the flare QPPs \citep[see][for a recent
review]{Zimovets21}. They could be regarded as a type of MHD
wave, such as the slow-mode wave, the kink-mode wave, or the
sausage-mode wave \citep[see][for
reviews]{Lib20,Nakariakov21,Wang21}. This mechanism can be used to
explain those flare QPPs that are strongly associated with magnetic
loops and current sheets \citep{Zimovets21}. The flare QPPs could be
driven by the repetitive regime of magnetic reconnection, which
would periodically accelerate non-thermal ions and electrons
\citep{Thurgood17,Karampelas22,Corchado24}. The quasi-periodic
reconnection may be spontaneous such as the self-oscillatory process
\citep{McLaughlin09} and the magnetic tuning fork model
\citep{Takasao16}, or might be triggered by an external wave
\citep{Kumar16,Nakariakov18,Li22}. These models well explain
those QPPs seen in HXR and microwave emissions during the flare
impulsive phase \citep[e.g.,][]{Yuan19,Clarke21,Luo22,Li22b,Li23}.
Besides, the flare QPPs are also modulated by the LRC-circuit
oscillation, which needs the current-carrying plasma loop
\citep[e.g.,][]{Tan16,Li20b}. It appears that one mechanism could
well trigger some certain flare QPPs, but it failed to drive all
the observed QPPs, and the generation mechanism seems to be related
to the observed period or the various wavebands
\citep[e.g.,][]{Kupriyanova20}.

Recently, the localizing QPPs were simultaneously observed in HXR
and microwave emissions, but the measured quasi-periods were always
shorter than 60~s \citep[e.g.,][]{Kou22,Collier24,Shi24}. In this
study, we localize the flare QPPs at a long period of about 200~s,
which are simultaneously seen in wavelengths of HXR, microwave, and
Ly$\alpha$ during an X-class flare. The article is organized as
follows: Section~2 introduces the observations, Section~3 shows the
data analysis and our main results, and Section~4 presents the
conclusion and discussion.

\section{Observations}
We analyze an X6.4 flare that occurred in the active region of
NOAA~13590 on 2024 February 22. It began at $\sim$22:08~UT, reached
its SXR maximum at $\sim$22:34~UT, and ended at about
$\sim$22:43~UT\footnote{https://www.solarmonitor.org/?date=20240223}.
The X6.4 flare was simultaneously measured by the Hard X-ray Imager
\citep[HXI;][]{Su19,Zhang19} and the Lyman-alpha Solar Telescope
\citep[LST;][]{Chen19,Feng19} on board the Advanced Space-based Solar
Observatory \citep[ASO-S;][]{Huang19,Gan23}, the
Spectrometer/Telescope for Imaging X-rays \citep[STIX;][]{Krucker20}
on Solar Orbiter, the Solar Upper Transition Region Imager
\citep[SUTRI;][]{Bai23}, the Geostationary Operational Environmental
Satellite (GOES), the Atmospheric Imaging Assembly
\citep[AIA;][]{Lemen12} and the Helioseismic and Magnetic Imager
\citep[HMI;][]{Schou12} aboard the Solar Dynamics Observatory (SDO),
the Nobeyama Radio Polarimeters (NoRP), and the Expanded Owens
Valley Solar Array \citep[EOVSA;][]{Gary11}.

HXI provides the HXR imaging spectroscopy of solar flares in the
energy range of $\sim$15$-$300~keV. The time cadence is 4~s in the
regular mode and could be as high as 0.125~s in the flare mode. We
use the data product at Level~1 (L1) in the HXR energy range of
20$-$300~keV, both the light curves and images are analyzed. STIX
measures X-ray imaging spectroscopy in the energy range of
4$-$150~keV during solar flares. The data product of QL1 at
25$-$84~keV is used, which has a time cadence of 4~s. The advantage
of this data type is that we can use it as early as possible since
it releases quickly. GOES provides SXR light curves at 1$-$8~{\AA}
and 0.5$-$4~{\AA} with a time cadence of 1~s. NoRP measures solar
microwave fluxes at six frequencies with a time cadence of 1~s.
EOVSA takes the solar radio dynamic spectrum in the frequency range
of $\sim$1$-$18~GHz, and its time cadence can be as high as 1~s, but
some data gaps occasionally appear in the EOVSA spectrum.

AIA provides full-disk solar maps in EUV/UV wavebands, the time
cadence of seven EUV wavebands is 12~s, and that of two UV wavebands
is 24~s. In this study, we use AIA maps in seven wavebands of
193~{\AA} ($\sim$20~MK), 131~{\AA} ($\sim$11~MK), 94~{\AA}
($\sim$6.3~MK), 171~{\AA} ($\sim$0.63~MK), 1600~{\AA}
($\sim$0.1~MK), 304~{\AA} ($\sim$0.05~MK), and 1700~{\AA}
($\sim$0.005~MK). We want to state that the temperature in each
waveband refers to the flare model. HMI takes the full-disk solar
magnetogram, dopplergram, and continuum filtergram. We analyze the
line-of-sight (LOS) magnetogram, which has a time cadence of 45~s.
The AIA maps and HMI magnetogram have been pre-processed by
``aia\_prep.pro'' and ``hmi\_prep.pro'', and thus they both have a
pixel scale of 0.6\arcsec. LST includes two telescopes, the Solar
Disk Imager (SDI) takes the Ly$\alpha$ maps at 1216~{\AA}, and the
White-light Solar Telescope captures white-light maps at 3600~{\AA}.
Similar to HXI, the cadence of SDI is 60~s in the regular mode and
changes to $\sim$6~s in the burst mode. In this study, the SDI
images in the regular mode is used, since the burst mode misses some
data during the impulsive phase of the X-class flare. SUTRI takes
full-disk solar maps at Ne~VII~465~{\AA} \citep{Tian17}. It has a
time cadence of roughly 30~s and a pixel scale of $\sim$1.23\arcsec.
However, only a few maps are captured by SUTRI during the X6.4
flare, which is used to show the flare shape at 465~{\AA}.

\section{Data Analysis and Results}
\begin{figure}
\centering
\includegraphics[width=\linewidth,clip=]{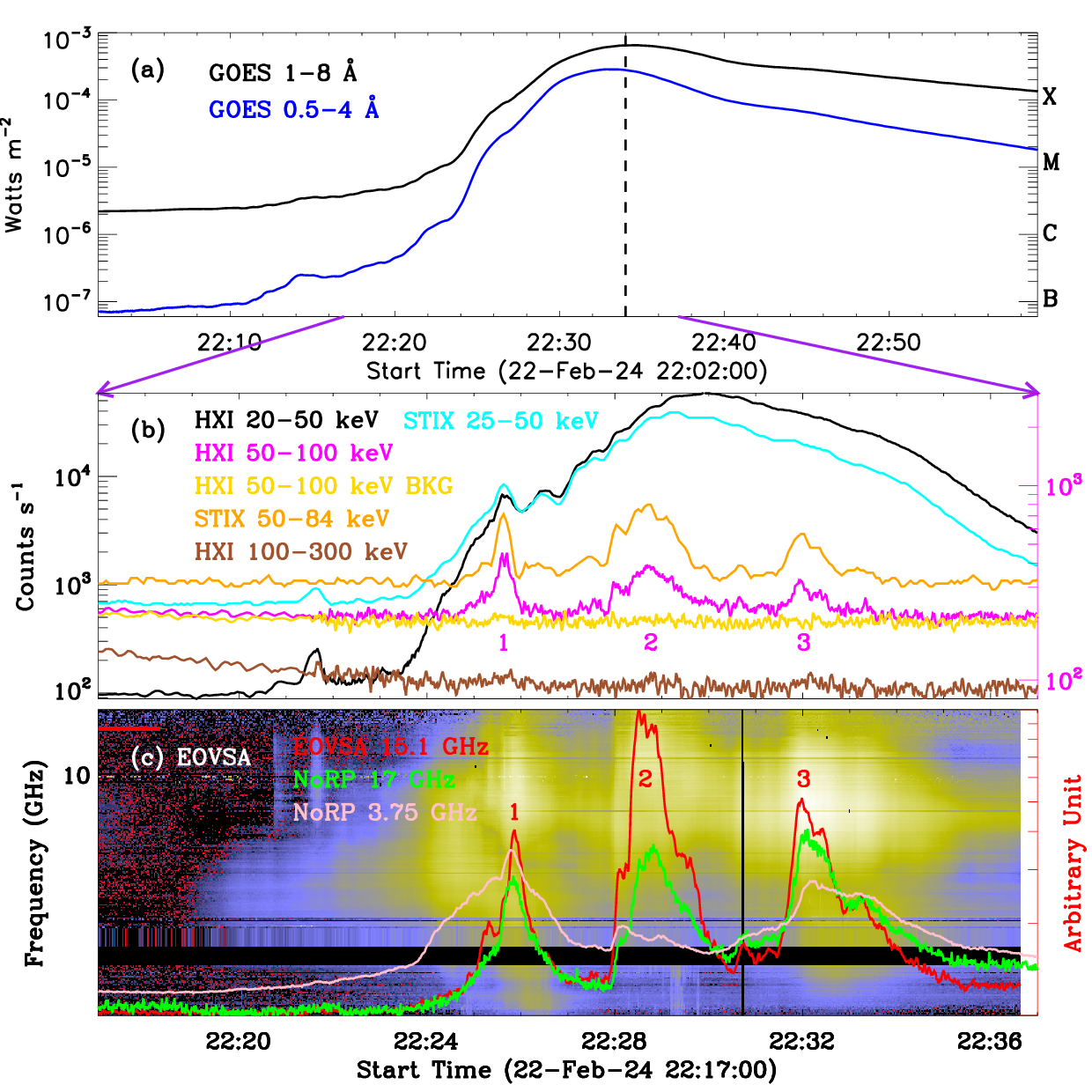}
\caption{Full-disk light curves of the X6.4 flare on 2024 February
2. (a) SXR light curves measured by GOES~1$-$8~{\AA} (black) and
0.5$-$4~{\AA} (blue). The vertical dashed line indicates the peak
time at GOES~1$-$8~{\AA}. (b) HXR fluxes recorded by HXI~20$-$50~keV
(black), 50$-$100~keV (magenta) and its background (gold), and
100$-$300~keV (brown), as well as STIX~25$-$50~keV (cyan) and
50$-$84~keV (orange). (c) Microwave fluxes captured by
EOVSA~15.1~GHz (red), and NoRP~17~GHz (green) and 3.75~GHz (pink).
The context image is the radio dynamic spectrum observed by EOVSA.
\label{over}}
\end{figure}

Figure~\ref{over} presents the full-disk light curves in wavelengths
of SXR/HXR and microwave. Panel~(a) shows the SXR fluxes recorded by
GOES at 1$-$8~{\AA} and 0.5$-$4~{\AA} from 22:02~UT to 22:59~UT on
2024 February 22, which shows an X6.4-class flare and reaches its
maximum at $\sim$22:34~UT in GOES~1$-$8~{\AA}, as indicated by the
dashed vertical line. Panel~(b) draws the HXR fluxes during
22:17$-$22:37~UT measured by ASO-S/HXI at 20$-$50~keV, 50$-$100~keV
and its background (BKG), and 100$-$300~keV, as well as captured by
STIX at 25$-$50~keV and 50$-$84~keV. Noting that STIX was located at
0.74~AU, and it was 26.5$^{\circ}$ from the Sun-Earth line in this
event. One can immediately notice that three successive pulsating
features simultaneously appear in the light curves of
HXI~50$-$100~keV and STIX~50$-$84~keV between about 22:24~UT and
22:34~UT, as marked by the Arabic numerals `1', `2' and `3'. The
three successive pulsations might be regarded as a candidate of
flares QPPs in the HXR channel. Conversely, we do not find
similar three pulsations in the HXR fluxes at HXI~20$-$50~keV and
100$-$300~keV, and STIX~25$-$50~keV. The light curves at
HXI~20$-$50~keV and STIX~25$-$50~keV seem to reveal a group of
pulses with a small amplitude, which is absolutely different from
those three successive pulsations with a large amplitude. There is
not any apparent signature in the light curve at HXI~100$-$300~keV,
suggesting that the high-energy particles are rarely accelerated in
the X6.4 flare. Panel~(c) shows the microwave light curves and the
EOVSA spectrogram. Similar to those HXR fluxes at HXI~50$-$100~keV
and STIX~50$-$84~keV, these microwave fluxes in the high frequencies
of NoRP~17~GHz and EOVSA~15.1~GHz also show three successive
pulsations with a large amplitude during $\sim$22:24$-$22:34~UT,
which could be considered as the QPP feature in the microwave
emission. On the other hand, the microwave flux in the low microwave
frequency of NoRP~3.75~GHz appears to show more than three
pulsations, and their durations are obviously longer than those seen
in the high microwave frequency. The radio dynamic spectrum in the
frequency range of $\sim$1$-$18~GHz firmly confirms that the
microwave radiation at the low frequency has longer pulses. So, we
do not analyze the microwave flux at the low frequency. It should be
pointed out that the black vertical line in the EOVSA spectrogram is
mainly due to the data gap.

\begin{figure}
\centering
\includegraphics[width=\linewidth,clip=]{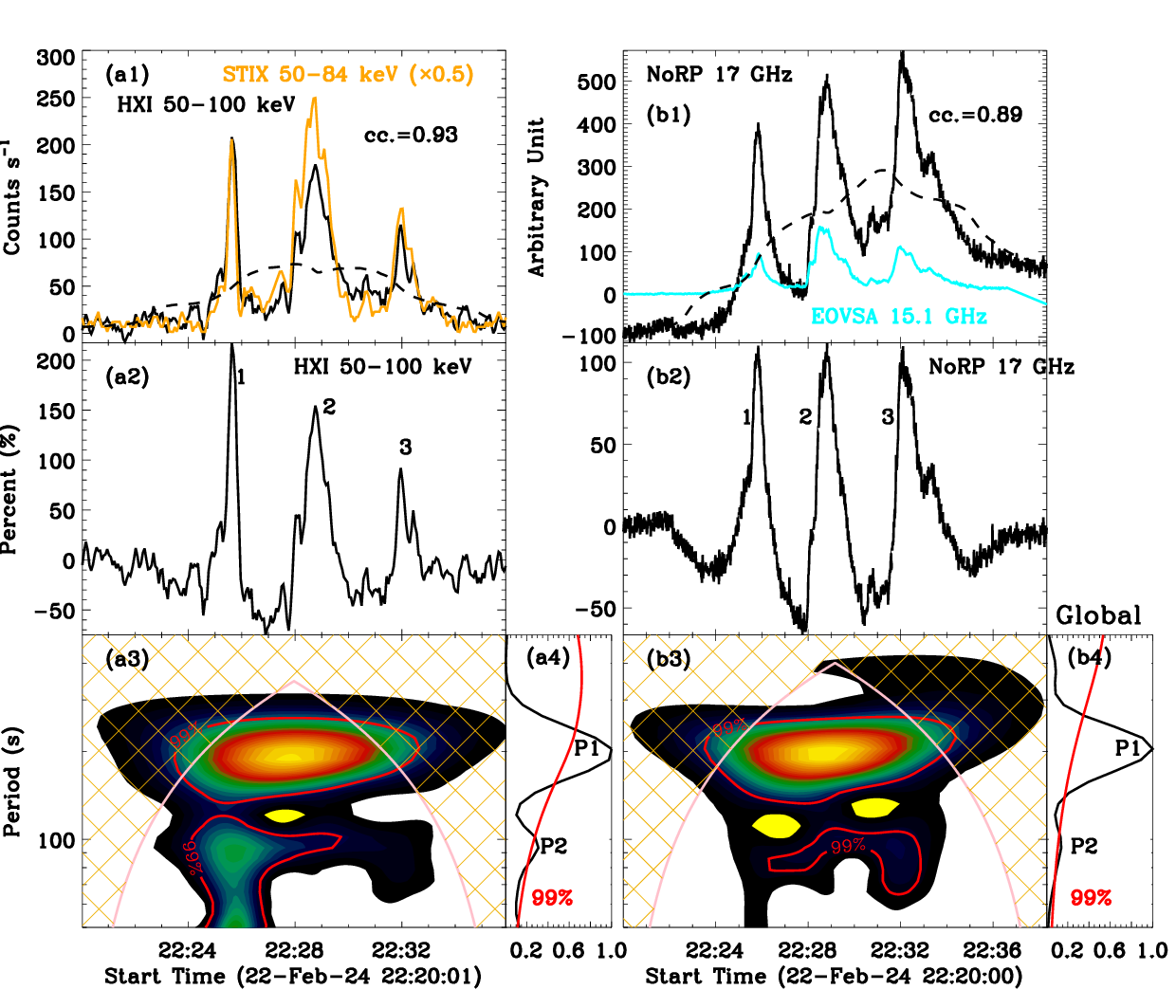}
\caption{Morlet wavelet analysis results for the HXR and microwave
light curves. (a1) \& (b1) HXR and microwave fluxes recorded by
HXI~50$-$100~keV, STIX~50$-$84~keV, NoRP~17~GHz, and EOVSA~15.1~GHz,
respectively. Here, the HXI flux has been subtracted from its background
emission. The overlaid dashed lines represent long-term trends at
HXI~50$-$100~keV and NoRP~17~GHz. (a2) \& (b2) Detrended light
curves normalized to their maximum of long-term trends in
wavelengths of HXI~50$-$100~keV and NoRP~17~GHz. (a3) \& (b3) Morlet
wavelet power spectra. (a4) \& (b4) Global wavelet power. The red
contours or lines represent the significance level of 99\%.
\label{wav1}}
\end{figure}

In order to examine the periodicity of the flare QPP, we applied the
wavelet transform method within a mother function of `Morlet'
\citep{Torrence98}, as shown in Figure~\ref{wav1}. Here, the HXR
light curve recorded by HXI is interpolated as a uniform
cadence of 4~s. Panels~(a1)$-$(a4) present the Morlet wavelet
analysis results in the HXR channel. The raw light curves recorded
by HXI~50$-$100~keV and STIX~50$-$84~keV appear to correlate
remarkably well with each other, and their linear Pearson
correlation coefficient (cc.) is about 0.93, as shown in panel~(a1).
Noting that the HXI flux at 50$-$100~keV has been removed its
background emission, and the long-term trend at HXI~50$-$100~keV is
calculated by a smoothing window of 6~minutes, as indicated by the
dashed curve. The 6-minutes window is chosen because we want
to suppress the long-period trend and enhance the short-period QPP,
i.e., the quasi-period of 3$-$4~minutes
\citep[e.g.,][]{Yuan19,Li20a,Li22b}. Panel~(a2) shows the detrended
time series at HXI~50$-$100~keV, which is normalized to the maximum
of its long-term trend. The detrended time series also reveals three
successive pulsations, and they appear to a one-to-one
correspondence with those pulsations in the raw HXR flux, confirming
that the smoothing window can enhance the QPP feature. Moreover, the
modulation depth of the QPP feature, which is regarded as the ratio
between the oscillating amplitude and the maximum of its long-term
trend, can reach roughly 210\%. Then, the wavelet power and
global wavelet power spectra of the detrended time series are shown
in Figure~\ref{wav1}~(a3) and (a4). They are mainly dominated by two
bulks of power spectra inside the 99\% significance level, implying
the presence of double periods. The dominant periods are identified
from the peak/center of the global wavelet power spectrum, and their
uncertainties could be determined by the half full width at the 99\%
significance level. So, double periods of $\sim$200$\pm$40~s and
$\sim$95$\pm$10~s are identified in the HXR channel.

Figure~\ref{wav1}~(b1)$-$(b4) show the Morlet wavelet analysis
results in the microwave emission. The raw light curves measured by
NoRP~17~GHz and EOVSA~15.1~GHZ agree well with each other, and their
correlation coefficient is about 0.89, as shown in
panel~(b1). The overlaid dashed curve is the long-term trend with the
same smoothing window of 6-minutes at NoRP~17~GHz. Panel~(b2) draws the
normalized detrended time series, as normalization by the maximum of
its long-term trend. It shows similar three
successive pulsations and the modulation depth of the QPP
feature could be as high as 110\%. Panels~(b3) and (b4) present
the wavelet power and global wavelet power spectra for the detrended
time series at NoRP~17~GHz. Similar to what has been seen in the HXR
channel, double periods within large uncertainties are determined
from the global wavelet power spectrum, which are $\sim$200$\pm$50~s
and $\sim$95$\pm$10~s.

\begin{figure}
\centering
\includegraphics[width=\linewidth,clip=]{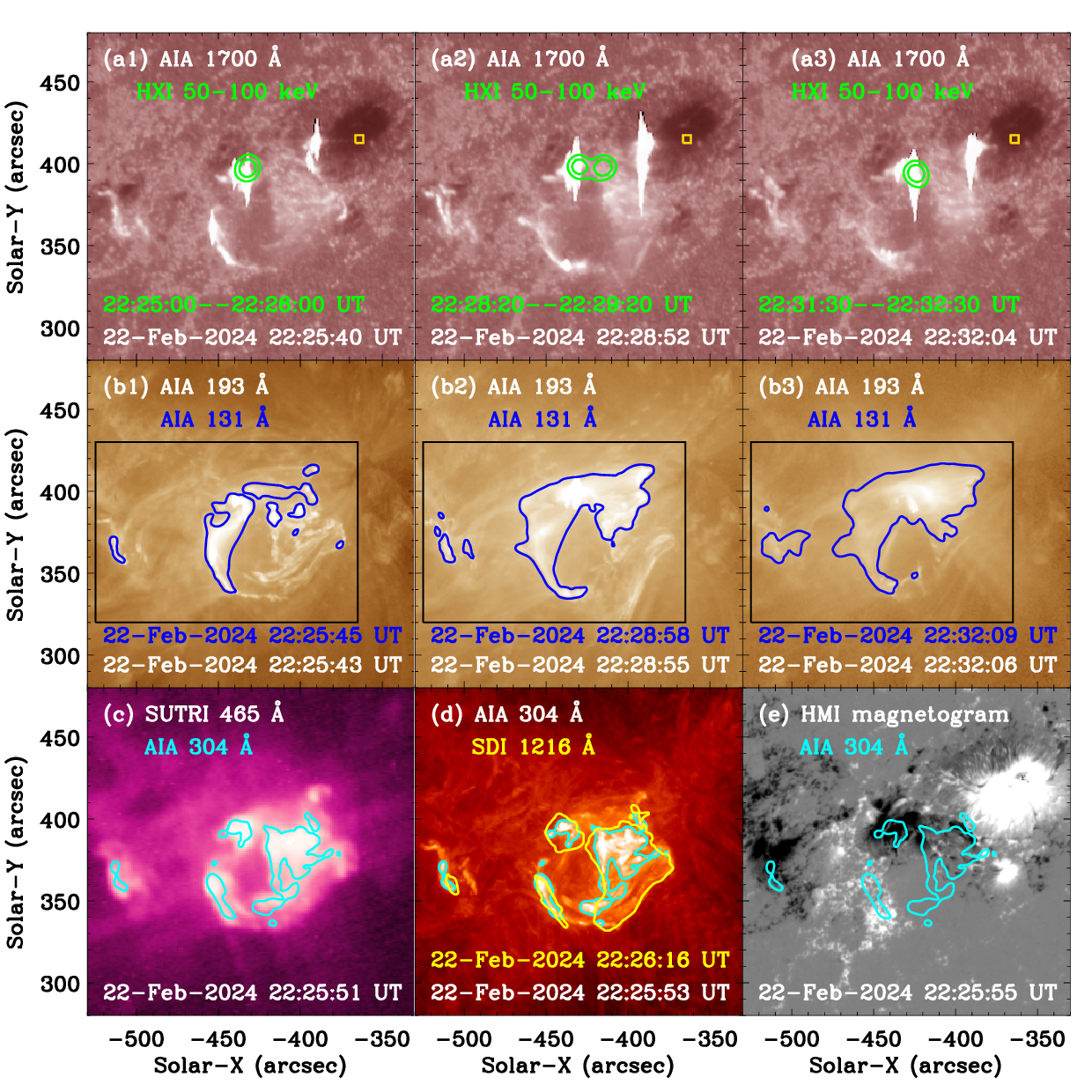}
\caption{Multi-wavelength snapshots with a FOV of
$\sim$200\arcsec$\times$200\arcsec\ during the X6.4 flare.
(a1)$-$(a3) UV maps measured by AIA~1700~{\AA}. The green contours
represent the HXR radiation observed by ASO-S/HXI at 50$-$100~keV,
and the levels are set at 20\% and 50\%, respectively. The gold box
outlines an umbral region, which is applied to integrate over the
umbral light curve. (b1)$-$(b3) EUV maps measured by AIA~193~{\AA},
and the blue contours are the AIA~131~{\AA} emissions. The black
rectangle outlines the flare area used to integrate over the flare
fluxes. (c) \& (d) EUV maps captured by SUTRI~465~{\AA} and
AIA~304~{\AA}. The cyan and yellow contours are the EUV emissions at
AIA~304~{\AA} and SDI~1216~{\AA}, respectively. (e) HMI LOS
magnetogram with a strength scale of $\pm$800~G. \label{img}}
\end{figure}

To localize the generated source of these long-periodic pulsations,
Figure~\ref{img} shows the multiple-wavelength images during the
three pulsations. Panels~(a1)$-$(a3) plot the UV maps with a
field-of-view (FOV) of $\sim$200\arcsec$\times$200\arcsec\ in the
wavelength of AIA~1700~{\AA} at three time instances. It can be seen
that the X6.4 flare is located near a sunspot and exhibits some bright
features in those UV maps. The main body of the X6.4 flare
consists of several kernel-like structures, and they form a
quasi-circular profile shape. A remote ribbon-like structure can be
seen in the east of the main body. All those observational facts
suggest that the X6.4 flare may be regarded as a circular-ribbon
flare. The HXR source in the energy range of HXI~50$-$100~keV is
overlaid on the brightest kernel, as outlined by the green contours,
and the contour levels are set at 20\% and 50\%, respectively. Here,
the HXR images during three pulsations are reconstructed by the
HXI\_CLEAN method, utilizing the detectors from D29 to D91, i.e.,
the subcollimator group G4 to G10. The fine grids of G1 to G3 are
excluded, because they are not calibrated well and we do not focus
on the fine structure \citep[cf.][]{Li23}. The gold box outlines an
umbral region in the sunspot, which is used to integrate the umbral
light curve. Figure~\ref{img}~(b1)$-$(b3) show the EUV maps in the
high-temperature channel of AIA~193~{\AA} and 131~{\AA} (blue
contours). We can see that the X6.4 flare shows several loop-like
structures that connect those kernel-like structures seen in
AIA~1700~{\AA} maps, it also shows a remote loop-like feature. The
black rectangle outlines the flare area used to integrate the local
light curves. Panels~(c) and (d) present the EUV and Ly$\alpha$ maps
in wavelengths of SUTRI~465~{\AA}, AIA~304~{\AA} (cyan contours),
and SDI~1216~{\AA} (yellow contours), and their formation
temperature are much lower than that in AIA~193~{\AA} and 131~{\AA}.
They manifested as a main circular profile shape and a remote ribbon
structure, which can be used to co-align those maps measured by
various instruments, as indicated by the cyan and yellow contours.
Figure~\ref{img}~(e) shows the LOS magnetogram observed by HMI, and
it shows that the main circular-ribbon feature is located in
positive and negative magnetic fields, which formed a quasi-circular
shape, while the remote ribbon appears in the negative magnetic
field. All those imaging observations suggest that the X6.4 flare is
a circular-ribbon flare with a fan-spine topology, which is produced
in the complicated magnetic structure.

\begin{figure}
\centering
\includegraphics[width=\linewidth,clip=]{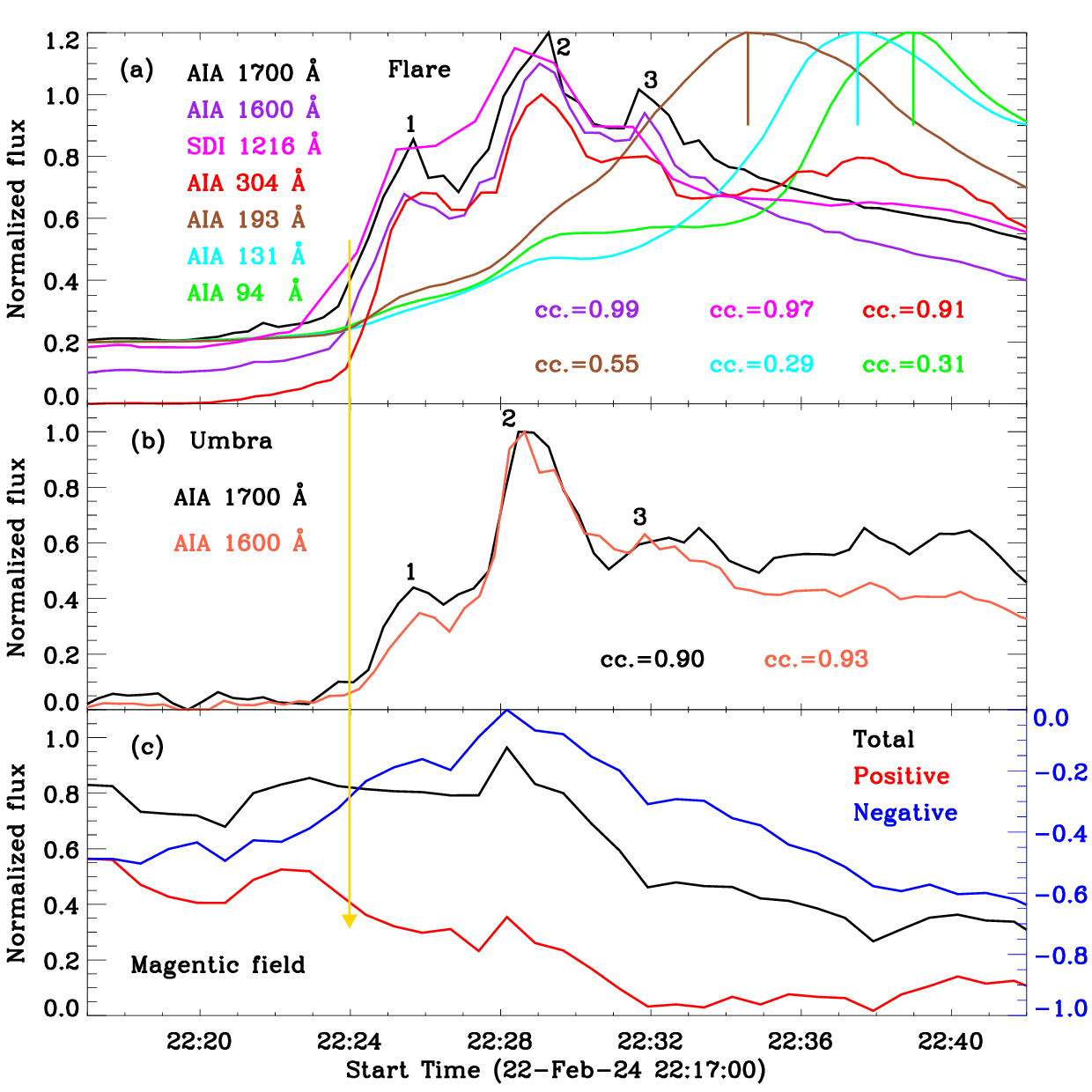}
\caption{Local light curves during the X6.4 flare. (a) UV/EUV and
Ly$\alpha$ fluxes integrated over the flare area in wavelengths of
AIA~1700~{\AA} (black), 1600~{\AA} (purple), 304~{\AA} (red),
193~{\AA} (brown), 131~{\AA} (cyan), and 94~{\AA} (green), and
SDI~1216~{\AA} (magenta). The numbers mark three pulsations in the
low-temperature channels, and the vertical short lines indicate the
peak time of the high-temperature channels. (b) UV light curves
integrated over the umbral region in wavelengths of AIA~1700~{\AA}
(black) and 1600~{\AA} (tomato). (c) Time series of the magnetic
flux integrated over the flare area for the total, positive, and
negative magnetic fields, respectively. The gold arrow outlines the
onset time of flare QPPs and the magnetic cancellation. \label{flux}}
\end{figure}

Figure~\ref{flux}~(a) presents the local light curves integrated
over the flare area in multi-wavelengths measured by SDO/AIA and
ASO-S/LST/SDI. In this study, the time cadence of SDO/AIA is chosen
as 24~s for all AIA maps to avoid imaging saturation
\citep[cf.][]{Li15}. These light curves are used as normalization,
i.e., $\frac{I-I_{\rm min}}{I_{\rm max}-I_{\rm min}}$, where $I$
represents the measured intensity, $I_{\rm min}$ and $I_{\rm max}$
are the minimum and maximum intensities. Similar to what has been
seen in the HXR and microwave fluxes, the UV/EUV and Ly$\alpha$
light curves at low-temperature channels (e.g., AIA~1700~{\AA},
1600~{\AA} and 304~{\AA}, SDI~1216~{\AA}) appear for three successive
pulsations during $\sim$22:24$-$22:34~UT, suggesting the presence of
flare QPPs with a quasi-period of roughly 200~s. On the other hand,
the EUV light curves at high-temperature channels (e.g.,
AIA~193~{\AA}, 131~{\AA}, 94~{\AA}) do not show any signature of the
QPP feature with the same period. It is interesting that the peak
time of the high-temperature channels is later and later with
a decrease in the formation temperature, as indicated by the
vertical short lines, suggesting that the hot flare loops are
cooling with time. We also calculate the linear Pearson correlation
coefficients (cc.) of these light curves, and they are all relative
to the light curve at AIA~1700~{\AA}. Obviously, the correlation
coefficients of those light curves in the low-temperature channels
are all larger than 0.9, while those in the high-temperature
channels are a bit small, i.e., $\leq$0.55. These observational
facts confirm that the QPP feature could only be seen in the
low-temperature channels.

We note that the X6.4 flare is located on the edge of a sunspot, and
the quasi-period of the flare QPP is roughly equal to 3~minutes.
Therefore, we also plot the local light curves integrated over the
umbral region in the neighboring sunspot at AIA~1700~{\AA} and
1600~{\AA}, as shown in Figure~\ref{flux}~(b). Similar to those
flare fluxes, they both show three successive pulsations, as marked
by the arabic numerals, suggesting the presence of similar periods.
Moreover, they have high correlations with the flare light curve at
AIA~1700~{\AA}, i.e., their correlation coefficient could be as high
as 0.90/0.93. At last, the normalized magnetic fluxes integrated
over the flare area are given in panel~(c), including the total,
positive, and negative magnetic fluxes. It can be seen that both the
positive and negative magnetic fluxes appear to decrease when the
start of flare QPPs, as indicated by the gold arrow. The total
magnetic flux reveals a significant reduction, indicating a process
of magnetic cancellation. These observations imply the occurrence of
magnetic reconnection during the flare QPPs. However, we do not see
similar three pulsations in the time series of
magnetic fluxes, which might be due to the lower sensitivity and
time resolution of SDO/HMI.

\begin{figure}
\centering
\includegraphics[width=\linewidth,clip=]{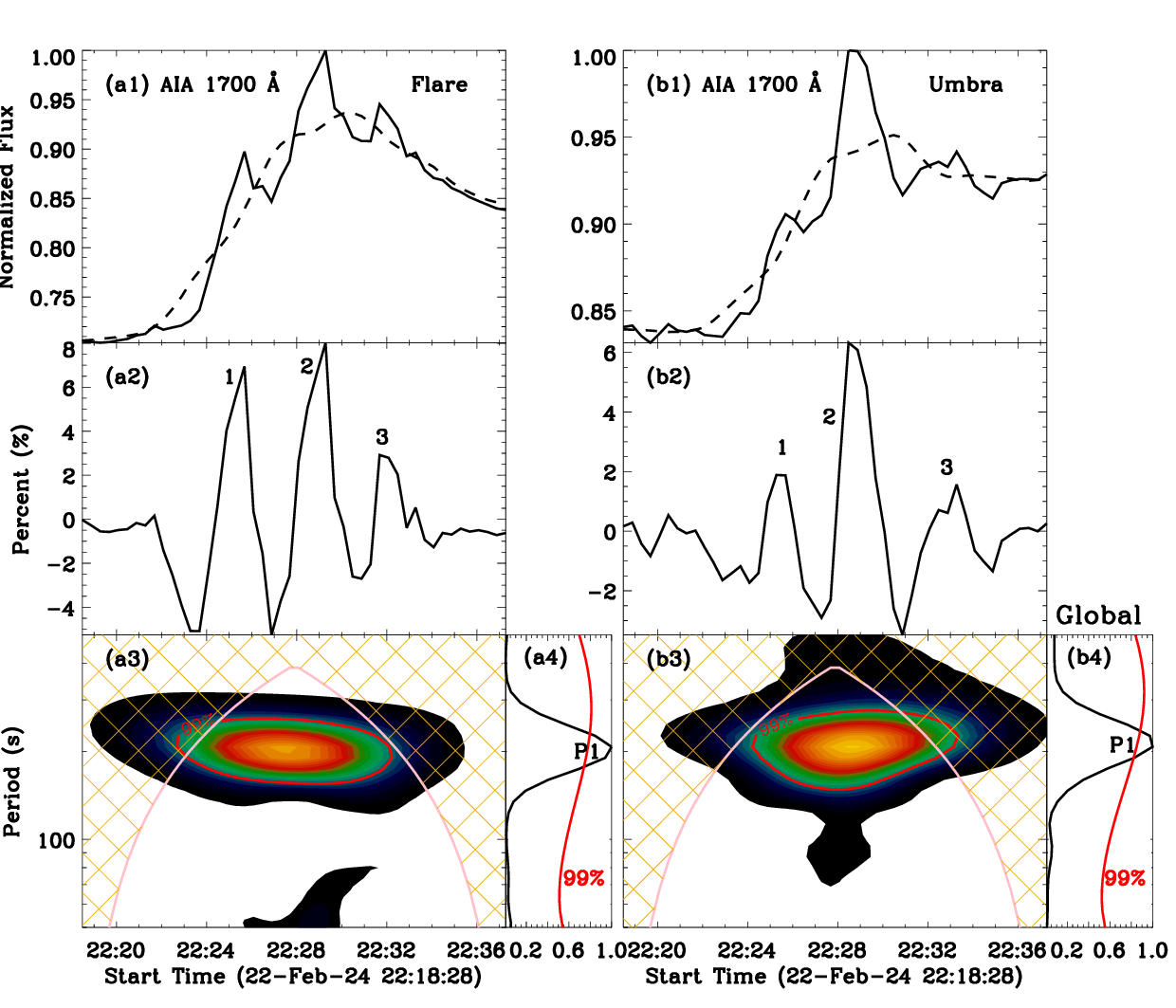}
\caption{Similar to Fig.~\ref{wav1} but the wavelet analysis is
performed for the AIA~1700~{\AA} light curves integrated over the
flare area (a1$-$a4) and the adjacent sunspot umbra (b1$-$b4),
respectively. \label{wav2}}
\end{figure}

Figure~\ref{wav2} shows the Morlet wavelet analysis results in the
wavelength of AIA~1700~{\AA} at the flare area (a1$-$a4) and the
umbral region (b1$-$b4), respectively. Using the same smoothing
window of 6~minutes, the raw light curve is decomposed into the
long-term trend (dashed lines) and the detrended time series (a2 and
b2). We should state that the raw light curves are normalized to
their maximum intensity, and the detrended time series are
normalized to the maximum of their long-term trends. Thus, we can
estimate the modulation depth of the QPP feature, which is
identified as the ratio between the oscillating amplitude and the
maximum of its long-term trend. The modulation depth is about
6\%$-$8\%, which is much lower than that observed in the
HXR/micorwave emission. The quasi-periods can be determined from the
wavelet power or global wavelet power spectra, which are estimated
to be $\sim$205$\pm$25~s in the flare area, and $\sim$205$\pm$20~s
in the umbral region, as shown in panels~(a4) and (b4). Here, the
short period at about 95~s is not detected, which is different from
what has been observed in the HXR/micorwave channel.

\begin{figure}
\centering
\includegraphics[width=\linewidth,clip=]{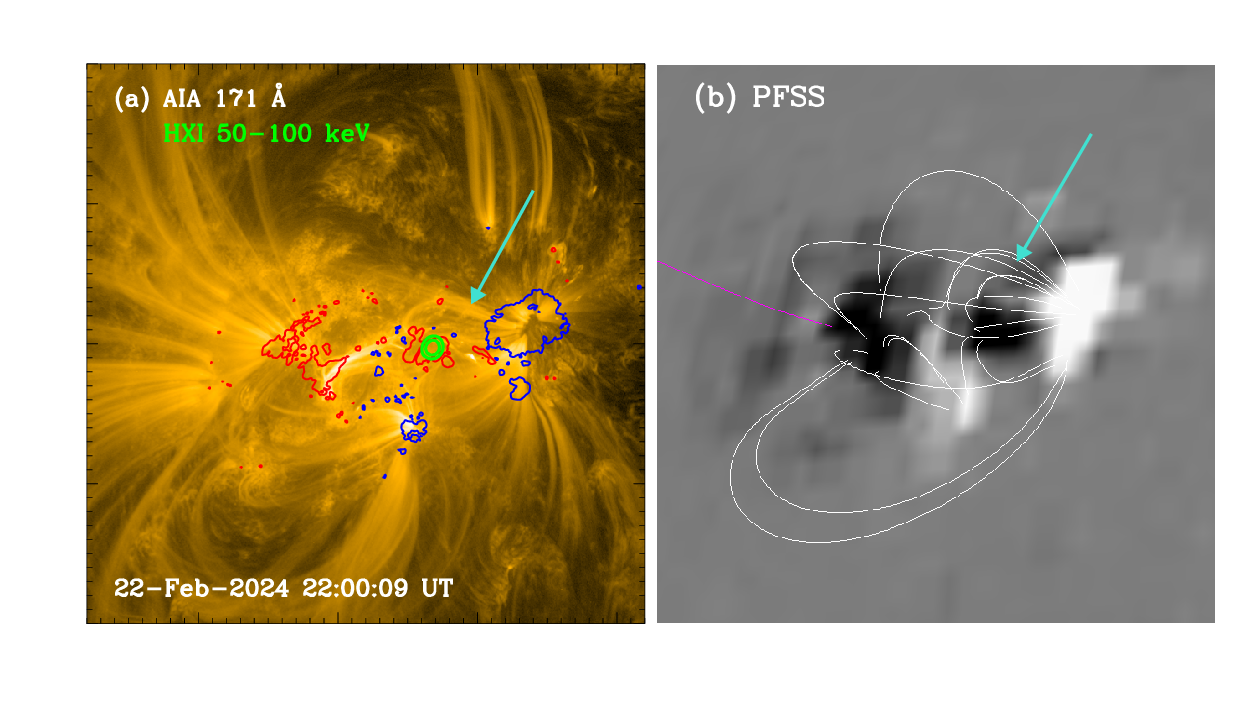}
\caption{The active region of NOAA 13590 before the initiation of
the X6.4 flare. (a) The EUV snapshot at AIA~171~{\AA}. The blue and
red contours are the positive and negative magnetic fields at the
strength scale of $\pm$500~G, respectively.  (b) The magnetic
configuration derived from a PFSS model. The white and purple lines
represent the closed and open magnetic fields. The turquoise arrow
indicates some coronal loops or magnetic field lines that connect
the flare core (green contours) and the sunspot. \label{loop}}
\end{figure}

In order to offer much more insight in the structure of
magnetic fields above the targeted active region, two maps showing
the active region of NOAA 13590 before the initiation of the X6.4
flare are drawn in Figure~\ref{loop}. Panel~(a) presents the EUV map
at 22:00:09~UT in the wavelength of AIA~171~{\AA}, and it clearly
shows a number of coronal loops in the targeted active region. The
over-plotted blue and red contours represent the polarity magnetic
fields measured by SDO/HMI before the onset of the the X6.4 flare,
and the green contours indicate the core area of the X6.4 flare
observed by ASO-S/HXI. It can be seen that the interested sunspot
was dominated by a strong positive magnetic field, as outlined by
the large blue contour. And the sunspot is located below some coronal
fan-like structures. The flare core and the coronal fan above the
sunspot are connected by a loop system, as indicated by the
turquoise arrow. Figure~\ref{loop}~(b) shows the magnetic field
topology on the flaring active region, which is derived from the
potential field source surface (PFSS) extrapolation
\citep{Schrijver03} based on the synoptic HMI magnetogram. The
flaring active region is mainly dominated by a large number of
closed magnetic field lines, while the open magnetic field line only
accounts for a small portion, as marked by the white and purple
lines. We also notice that the flare area and the sunspot are
connected by some closed field lines, as indicated by the turquoise
arrow. Both the EUV map at AIA~171~{\AA} and the PFSS extrapolated
result suggest that the X6.4 flare and the sunspot are mutually
influenced via some magnetic loops.

\section{Conclusion and Discussion}
By using the combined HXR, microwave, Ly$\alpha$, and UV/EUV
observations from HXI, STIX, EOVSA, NoRP, LST, SUTRI, HMI, and AIA,
we investigate the long-periodic pulsations during the impulsive
phase of an X6.4 flare and localize their source locations through
reconstructing the HXR map at HXI~50$-$100~keV. Double periods are
determined by the Morlet wavelet transform technique in HXR and
microwave wavebands. We also explore the generation mechanism of
flare QPPs, which could be related to the umbral oscillation in the
neighboring sunspot, because the quasi-period of flare QPPs matches
the periodicity of umbral oscillations in the nearby sunspot.

In our case, three successive pulsations are simultaneously observed
in HXR, microwave, Ly$\alpha$, UV, and low-temperature EUV
wavelengths, which could be regarded as flare QPPs. The wavelet
analysis results suggest that the dominant period of the flare QPP
is $\sim$200~s, and the period uncertainties are varied in different
wavelengths. The flare QPPs at long quasi-periods are frequently
detected in wavelengths of Ly$\alpha$, UV/EUV, and SXR
\citep[e.g.,][]{Kumar15,Tan16,Milligan17,Lid22}. However, such
long-period QPP are rarely localized in HXR and microwave emissions.
Based on the observation from EOVSA, the flare QPPs with varying
periods of $\sim$10$-$20~s and $\sim$30$-$60~s were seen in the
microwave emission \citep[cf.][]{Kou22}. Combined the observations
from STIX and EOVSA, the flare QPPs with fast variations on
timescales evolving from $\sim$7~s to $\sim$35~s were found in HXR
and microwave channels \cite[cf.][]{Collier24}. Using the
observations from HXI, STIX, and the Chashan Broadband Solar
millimeter spectrometer, the flare QPP with a period of $\sim$27~s
is simultaneously found in HXR and microwave channels \citep{Shi24}.
Obviously, those quasi-periods are much shorter than that is
observed in our case, although they localised and imaged the flare
QPPs in HXR or microwave channels. On the other hand, the flare QPPs
at the quasi-period of about 4~minutes were simultaneously detected
in wavebands of HXR and low-frequency radio emissions, but they
identified the source locations of flare QPPs by using the imaging
observations at UV/EUV wavelengths \citep[e.g.,][]{Li15}.
\cite{Kumar16} reported the 3-minutes QPPs and determined their
source region by using the HXR and microwave observations, and the
3-minutes period could be simultaneously found in the
HXR~25$-$50~keV, microwave frequencies of~2~GHz, 3.75~GHz, 9.4~GHz,
and 17~GHz, as well as metric and decimetric radio frequencies of
25$-$180~MHz, 245~MHz, and 610~MHz. Here, the long-periodic
pulsations at periods of $\sim$200~s can be observed in the HXR
channel above 50~keV, microwave frequencies above 10~GHz, but they
can not be seen at HXR wavebands below 50~keV and radio frequencies
below 10~GHz. Interestingly, the long-periodic pulsations can also
be detected in the wavelengths of Ly$\alpha$, UV and low-temperature
EUV. The modulation depth of long-periodic pulsations, which is
defined as the ratio between the pulsation amplitude and the maximum
of its long-term trend, is extremely large in HXR and microwave
channels. For instance, it can reach 210\% in the HXR channel and
could be as high as 110\% in the high-frequency microwave emission.
They are much larger than the modulation depth in
wavelengths of UV/EUV and Ly$\alpha$, i.e., $\leq$8\%. Our results
in HXRs and microwaves are much bigger than previous findings, i.e.,
the modulation depth of $\sim$20\%$-$25\% in HXRs, $\gamma$-rays,
and microwave emission \citep[e.g.,][]{Nakariakov10,Li22}. Thus, the
long-periodic pulsations seen in the HXR and microwave channels show
an extremely-large modulation depth. This feature is
reasonable. Because the solar radiation in HXRs and $\gamma$-rays
are mainly from the flare area, their background emission is
quite weak, and the modulation depth could be large. Conversely, the
solar radiation in SXRs and UV/EUV can be from the entire Sun,
regardless of whether there is a flare occurring or not, so their
background emission is strong, resulting in the small-amplitude
oscillation
\citep[e.g.,][]{Simoes15,Ning17,Dominique18,Li20c,Shi24}.

Until now, the generation mechanism of flare QPPs is still not fully
understood \citep{Kupriyanova20,Zimovets21}. The long-periodic
pulsations are simultaneously observed in wavebands of HXR,
microwave, Ly$\alpha$, UV and low-temperature EUV during the flare
impulsive phase. Moreover, the magnetic cancellation occurs when the
onset of flare QPPs. Thus, they could be generated by nonthermal
electrons that are periodically accelerated by repetitive magnetic
reconnections \citep{Yuan19,Clarke21,Li23}. The HXR images show that
the source locations of these long-periodic pulsations are nearly
not moving over time, indicating that the repetitive magnetic
reconnections are not slipping \citep{Lit15}. Now the question is
whether the periodic reconnection is spontaneous or induced?
Considering the large uncertainty, the quasi-period of $\sim$200~s
is close to that of $\sim$3~minutes in the sunspot, particularly the
umbral oscillation in the chromosphere
\citep[e.g.,][]{Sych09,Kumar16,Li20d,Wang24}. By checking the
imaging observations captured by HXI and AIA, we find that the X6.4
flare is located at the edge of a sunspot. Then, a small umbral region
in the sunspot is chosen to look at its periodicity. Similar to the
flare area, the time series at the umbral region reveals a similar
quasi-period. Therefore, the long-periodic pulsations could be
associated with the umbral oscillation at the neighboring sunspot,
that is, it is triggered by the slow magnetoacoustic wave leaking
from the sunspot umbra in the photosphere, and then propagates
upwardly to the chromosphere and corona \citep[cf.][]{Yuan11}.
Both the AIA~171~{\AA} map and the PFSS extrapolation have
demonstrated that the flare core and the sunspot are linked by some
magnetic field lines, implying that the slow magnetoacoustic wave
could propagate from the coronal plasma fan above the sunspot to the
flare area along the magnetic loop \citep{Meadowcroft24}. Briefly,
the quasi-periodicity of magnetic reconnection during the X6.4 flare
is most likely to be induced by the slow magnetoacoustic wave
originating from the sunspot umbra, and it further modulates the
flare QPPs in wavelengths of HXR, microwave, Ly$\alpha$, UV and
low-temperature EUV during the impulsive phase
\citep{Clarke21,Zimovets21b,Li22}.

Another interesting point is that double periods are nearly
simultaneously observed in wavebands of HXR and microwave during the
impulsive phase of the X6.4 flare. The long period peaked at
$\sim$200~s is simultaneously seen in wavelengths of
HXI~50$-$100~keV, STIX~50$-$84~keV, EOVSA~15.1~GHz, NoRP~17~GHz,
SDI~1216~{\AA}, AIA~1600~{\AA}, 1700~{\AA}, and 304~{\AA}, while the
short period centered at $\sim$95~s can be only found in HXR and
microwave channels during the flare impulsive phase. The short
period is not observed in Ly$\alpha$ and UV wavelengths, largely due
to the low time resolution of LST/SDI and SDO/AIA. The ratio of
double periods are roughly equal to 2.1, which is consistent with
previous observations about the flare QPPs within double periods
\citep[e.g.,][]{Inglis09,Lu21,Lid22}. Our observational result
agrees with the theoretical expectation of the fundamental and
harmonic modes of weakly dispersive MHD waves \citep{Nakariakov05}.
So, the long period is regarded as the fundamental mode of the slow
magnetoacoustic wave that originated from the umbral oscillation
in the nearby sunspot, and the short period could be considered as
its second-harmonic mode.

\acknowledgments This work is supported by the National Key R\&D Program of China 2022YFF0503002 (2022YFF0503000), NSFC under grants 12333010 and 12073081, and the Strategic Priority Research Program of the Chinese Academy of Sciences, grant No. XDB0560000. D.~Li is also supported by
the Surface Project of Jiangsu Province (BK20211402) and Yunnan Key
Laboratory of Solar Physics and Space Science under the number of
YNSPCC202207. We thank the teams
of ASO-S, SUTRI, GOES, SDO, NoRP, and EOVSA for their open data use
policy. ASO-S mission is supported by the Strategic Priority
Research Program on Space Science, the Chinese Academy of Sciences,
Grant No. XDA15320000. SUTRI is a collaborative project conducted by
the National Astronomical Observatories of CAS, Peking University,
Tongji University, Xi'an Institute of Optics and Precision Mechanics
of CAS and the Innovation Academy for Microsatellites of CAS. SDO is
NASA's first mission in the Living with a Star program. The STIX
instrument is an international collaboration between Switzerland,
Poland, France, Czech Republic, Germany, Austria, Ireland, and
Italy.


\end{document}